\documentclass[prb,superscriptaddress,showpacs,
               reprint]{revtex4-1}

\usepackage[utf8]{inputenc}
\usepackage{amsmath,amsfonts,amssymb}
\usepackage{pifont}
\usepackage{rotating}
\usepackage{graphicx}    
\usepackage{epstopdf}
\usepackage{color}
\usepackage[caption=false]{subfig}
\usepackage{soul}
\usepackage{enumerate}

\usepackage{hyperref}
\definecolor{dark-red}{rgb}{0.9,0.15,0.15}
\definecolor{dark-blue}{rgb}{0.15,0.15,0.4}
\definecolor{medium-blue}{rgb}{0,0,0.5}
\hypersetup{
    colorlinks, linkcolor={black},
    citecolor={dark-blue}, urlcolor={medium-blue}
}

\begin{document}

\title{Competing magnetic and spin gap-less semiconducting behaviour in fully compensated ferrimagnet CrVTiAl: Theory and Experiment}

\author{Y. Venkateswara}
\thanks{Y. Venkateswara and Sachin Gupta contributed equally}
\affiliation{Magnetic Materials Laboratory, Department of Physics, Indian Institute of Technology Bombay, Mumbai 400076, India}

\author{Sachin Gupta}
\thanks{Y. Venkateswara and Sachin Gupta contributed equally}
\affiliation{Magnetic Materials Laboratory, Department of Physics, Indian Institute of Technology Bombay, Mumbai 400076, India}
\affiliation{WPI-Advanced Institute for Materials Research (WPI-AIMR), Tohoku University, Sendai 980-8577, Japan}

\author{S. Shanmukharao Samatham}
\affiliation{Magnetic Materials Laboratory, Department of Physics, Indian Institute of Technology Bombay, Mumbai 400076, India}

\author{Manoj Raama Varma}
\affiliation{National Institute for Interdisciplinary Sciences and Technology (CSIR), Thiruvananthapuram, India;}

\author{Enamullah}
\affiliation{Department of Physics, Indian Institute of Technology Bombay, Mumbai 400076, India}

\author{K. G. Suresh}
\email{suresh@phy.iitb.ac.in}
\affiliation{Magnetic Materials Laboratory, Department of Physics, Indian Institute of Technology Bombay, Mumbai 400076, India}

\author{Aftab Alam}
\email{aftab@iitb.ac.in}
\affiliation{Department of Physics, Indian Institute of Technology Bombay, Mumbai 400076, India}


\begin{abstract}
	
We report the structural, magnetic and transport properties of polycrystalline CrVTiAl alloy along with first principles calculations. It crystallizes in the LiMgPdSn type structure with lattice parameter 6.14 \AA\ at room temperature. Absence of (111) peak along with the presence of a weak (200) peak indicates the antisite disorder of Al with Cr and V atoms. The magnetization measurements reveal a ferrimagnetic transition near 710 K and a coercive field of 100 Oe at 3 K. Very low moment and coercive field indicate fully compensated ferrimagnetism in the alloy. Temperature coefficient of resistivity is found to be negative, indicating a characteristic of semiconducting nature. Absence of exponential dependence of resistivity on temperature indicates a gapless/spin-gapless semiconducting behaviour. Electronic and magnetic properties of CrVTiAl for three possible crystallograpic configurations are studied theoretically. All the three configurations are found to be different forms of semiconductors. Ground state configuration is a fully compensated ferrimagnet with band gaps 0.58 eV and 0.30 eV for up and down spin bands respectively. The next higher energy configuration is also ferrimagnetic, but has spin-gapless semiconducting nature. The highest energy configuration corresponds to a non-magnetic gapless semiconductor. The energy differences among these configurations are quite small ($<$ 1 $\mathrm{mRy/atom}$) which hints that at finite temperatures, the alloy  exists in a disordered phase, which is a mixture of the three configurations. By taking into account the theoretical and the experimental findings, we conclude that CrVTiAl is a fully compensated ferrimagnet with predominantly spin-gapless semiconductor nature.

\end{abstract}


\date{\today}

\maketitle

\section{Introduction}
Spintronics is an emerging branch of electronics in which the spin degree of freedom is added to the charge degree of electron to realize many
advantages such as non-volatility, high processing speed, low power consumption, high storage density etc. over the conventional electronics.\cite{Wolf1488,spinrev1,C.Felser,T.grafcFelser,prinzspin,sarma2001new,sarma2000theoretical} The utilization of the spin degree of freedom i.e., in spintronic devices, can be found in spin diodes used in magnetic hard disks, read heads, magnetoresistive random access memory (MRAM), spin transistors, tunnel diodes, vortex oscillators etc. \cite{emerginspin,Wunderlich1801,awschalom2007challenges,zhao2015spintronics,hu2011high} For realization of spintronic devices, special materials are required, for example, their electrical conduction should be restricted to one type of spin carriers. Such a phenomenon is seen in half metallic ferromagnets (HMF), spin gap-less semiconductors (SGS), semiconducting spin filters etc. \cite{deGrootNiMnSb,WandXL,Ouardiphyrevlett,MSandSGSgalanakis,C.Felser} Among the discovered materials, fully compensated ferrimagnetic (FCF) materials have gained a lot of interest recently.\cite{sgsreviewarticle1,ajaynaikprb1,v3alprb} Leuken and Groot showed theoretically  that this new class of materials can show 100 \% spin polarization without having a net magnetic moment, and hence given the name half metallic antiferromagnetic materials or fully compensated ferrimagnets. \cite{H.van} However, for the antiferromagnets, symmetry demands the same density of states (DOS) for spin up and spin down bands.\cite{v3alprb,MtasSasiglu2017} Due to symmetric bands and DOS, both the spin channels equally contribute to electrical conductivity, which results in zero net spin-polarized current. Such a scenario is not always true for FCF materials, which usually contain three or more magnetic ions with moments aligned in such a way that the net magnetization is nearly zero. Some of the unique properties and advantages of FCF materials  are (i) nearly zero magnetic moment which creates no external stray fields, resulting in low energy losses (ii) spin sensitivity without stray magnetic fields, which allows them not to disturb the spin character and make them ideal for spin polarized scanning tunnelling microscope tips and improved density of circuit integration in a chip \cite{spstm} (iii) low shape anisotropy, which helps in applications in spin injection etc. These properties make them superior compared to HMF  materials and are very much in demand today.
\begin{figure}[hbtp]
	\includegraphics[width=\linewidth, height=0.45\linewidth]{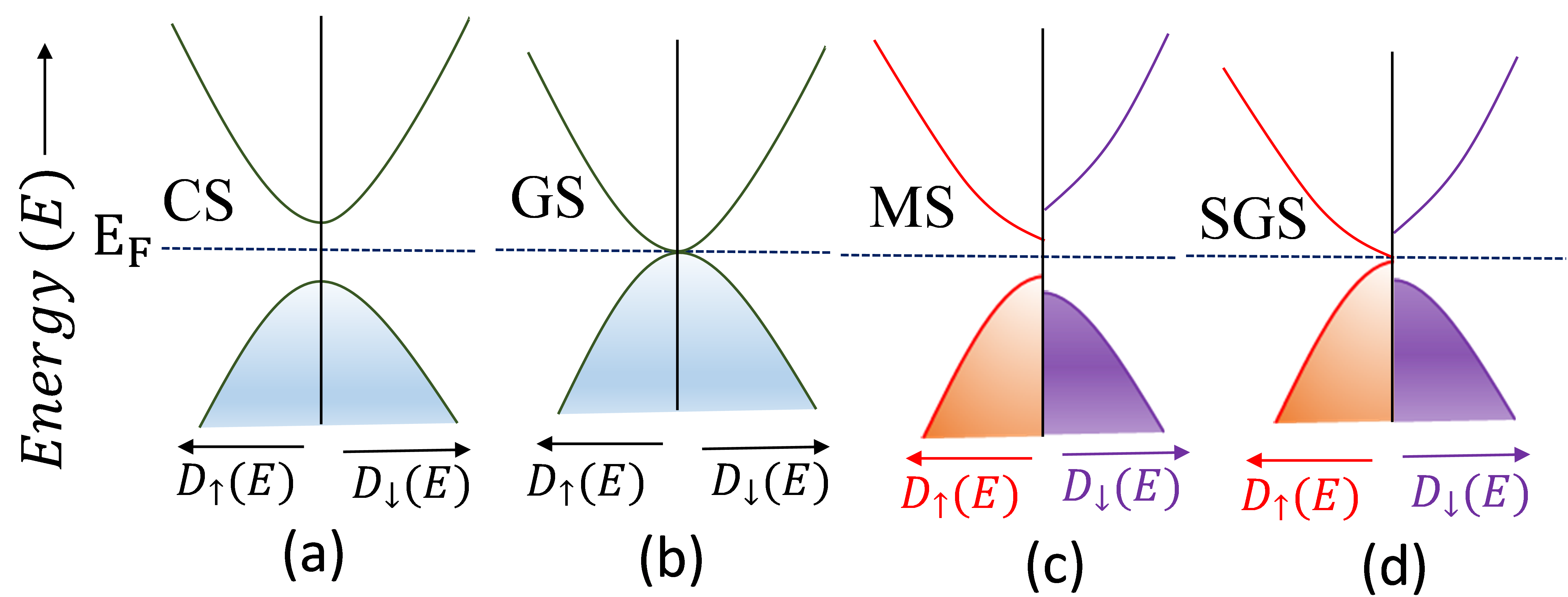} 
		\caption{Schematic density of states (DOS) of various types of semiconductors. a) conventional semiconductor (CS) in which both up ($\uparrow$) and down ($\downarrow$) spin bands have finite and equal band gaps. b) a gapless semiconductor (GS) where both the spin bands have vanishing band gap. c) a magnetic semiconductor (MS) in which the band gaps of up and down spin bands are finite but unequal. d) a spin-gapless semiconductor in which any one band (up or down) is gapless while the other is with finite gap. }
		\label{fig:schematic}
\end{figure}

Though Heusler alloys are known for many decades, they gained renewed interest because of the developments in the field of spintronics. \cite{T.grafcFelser,felser2013spintronics,fong2013half,felser2015heusler,sato2015spintronics,Warren_E} Ozdogan \text{et al.} studied the electronic and magnetic properties of quaternary Heusler alloys (QHAs) theoretically by using the full-potential non-orthogonal local-orbital minimum-basis band structure scheme (FPLO). \cite{k.Ozdogan-allQHA} Among the studied alloys, CrVTiAl (CVTA) has attracted a lot of interest and the preliminary band structure studies indicated it to be an antiferromagnetic semiconductor. Later, it was found that it is a fully compensated ferrimagnet with distinct magnetic moments at Cr, V and Ti ions. \cite{OzdoganCrVXAl,OzdoganCrVXZ} 

Schematic density of states of different classes of semiconductors are displayed in Fig. \ref{fig:schematic}. Figure \ref{fig:schematic} (a) is a conventional semiconductor (CS) in which both spin up and spin down bands have equal band gap ($\Delta E_g$). In thermal equilibrium, the intrinsic charge carrier concentration is given by \cite{kittel2007}

\begin{equation}
n_{i} = 2 \left( \dfrac{k_\mathrm{B}T}{2\pi \hbar^2} \right)^{3/2} (m_em_h)^{3/4} e^{-(\Delta E_g/2k_\mathrm{B}T)}.
\label{eq:intrinsic_density}
\end{equation}

Here $m_e$($m_h$) is the effective mass of electron(hole). The conductivity in CS is dominated by exponential term. Figure \ref{fig:schematic} (b) is the special case in which gap closes ($\Delta E_g \approx 0$) (for example $\mathrm{HgTe}$) in which $n_i$ varies as $T^{3/2}$.\cite{hgtebooksgs} Figure \ref{fig:schematic} (c) is a typical magnetic semiconductor in which band gap for each spin band is finite but not equal, resulting in spin polarized intrinsic carriers and hence used in spin filters. Figure \ref{fig:schematic}(d) is the special class of  magnetic semiconductors in which one of the spin bands encounters zero gap ($\Delta E_{g\uparrow} \approx 0$), while the other has a finite gap.\cite{WandXL,sgsmn2coalprb,Lakhancofemnsisgsprb} In this case, the concentration of intrinsic spin up carriers ($n_{\uparrow}$), which varies as $T^{3/2}$ dominates in comparison to that of spin down carries ({$n_{\downarrow}$}), which varies in an exponential manner and as a result, the temperature dependence of the total concentration of intrinsic carriers slightly deviates from pure $T^{3/2}$.

An experimental investigation carried out by Stephen \textit{et al.,} has shown CVTA to be a magnetic semiconductor,\cite{expcrvtialpaper} but they attributed the resistivity behaviour to a combination of metallic and semiconducting contributions. According to them, the magnetization depends linearly on the field, indicating the antiferromagnetic behaviour. However,  a close inspection of their XRD pattern reveals small peaks near (220), which is indicative of secondary phase(s). In addition, their sample shows a (111) peak with considerable intensity, unlike ours. The nearly equal electronegativities of Al and Cr/V causes the antisite disorder between these sites resulting the absence of superlattice (111) peak in the XRD.\cite{EYSprb} In view of these differences and with the aim of shedding more light into the anomalous properties exhibited by this alloy, we have carried out a combined theoretical and experimental study, which predicts entirely different set of properties than what is reported earlier.

\section{Experimental and theoretical details}

Polycrystalline CrVTiAl alloy was prepared by arc melting the stoichiometric proportions of constituent elements with purity at least 99.99 \%. Room temperature X-ray diffraction (XRD) patterns were collected using X’Pert Pro diffractometer using Cu K$\alpha$ radiation. The crystal structure was analyzed by Rietveld refinement using FullProf suite.\cite{fullprofRef} 

The crystal structure of QHAs of type $\mathrm{XX'YZ}$ (where X, $\mathrm{X'}$, Y are
transition elements and Z is the main group element), can be described by three distinct (symmetry inequivalent) possible arrangements of atoms. \cite{T.grafcFelser,EYSprb} The structure consists of 4 wyckoff sites 4a, 4b, 4c and 4d. By fixing Z at 4a site, the distinct configurations are 

\begin{enumerate}[(I)]
\item X at 4b, X$'$ at 4c and Y at 4d sites ,
\item X at 4c , X$'$ at 4b and Y at 4d sites, 
\item X at 4d , X$'$ at 4c and Y at 4b sites
\end{enumerate}
respectively. The structure factor for the first configuration is given by

\begin{equation}
F_{hkl}=4(f_z + f_y e^{\pi i(h+k+l)} + f_x e^{\frac{\pi i}{2} (h+k+l)} + f_{x'} e^{-\frac{\pi i}{2} (h+k+l)})
\end{equation}
with unmixed (hkl) values. Here $f_z$,$f_y$,$f_x$ and $f_{x'}$ are the atomic scattering factors for the atoms Z, Y, X and X$'$ respectively. Therefore, the magnitudes of 

\begin{eqnarray} 
F_{111}=4[(f_z-f_y)-i(f_x-f_{x'})]\\
F_{200}=4[(f_z+f_y)-(f_x+f_{x'})]\\
F_{220}=4[f_z+f_y+f_x+f_{x'}]
\label{eq:strufac}
\end{eqnarray}
are used to classify the order of the crystal structure.

Magnetization measurements (from 2-400 K) were performed implementing zero field cooled warming (ZFCW) and field cooled warming (FCW) protocols in 500 Oe using vibrating sample magnetometer (VSM, Quantum Design). High temperature magnetization measurement was carried out using (MPMS) in field warming (FW) mode at 1 kOe. Resistivity ($\rho$) measurements were carried out using Physical Property Measurement System (PPMS) by four probe method applying 5 mA current. 

\subsection{Theoretical Details}

Spin-resolved Density Functional Theory (SDFT) as implemented in Quantum Espresso (QE) package \cite{QEpack} was used to calculate the band structure and magnetic properties of CVTA. The exchange-correlation functional was taken within the generalized gradient approximation (GGA) in the parametrization of  Perdew-Burke-Ernzerhof (PBE).\cite{pbe} The pseudo potentials with Projector Augmented-Wave method \cite{kjpaw} were generated using PSlibrary and QE. Self consistent calculations were carried out using 24$\times$24$\times$24 k-point mesh with Methfessel-Paxton smearing of width 0.005 Ry, resulting in 413 $k$-points in the irreducible wedge of the Brillouin zone. The energy convergence criterion was set to $10^{-9}$ Ry. The kinetic energy of the plane wave expansion (energy cutoff $E_{cut}$) was restricted to 60 Ry and charge density expansion to 700 Ry. Non-self consistent field calculations were carried out using 48$\times$48$\times$48 k-point grid. Projected density of states (DOS) were extracted with an energy width of 0.0025 Ry.

Thermal charge carrier concentration was calculated using theoretical DOS, $D(E)$. The electron density above the Fermi energy ($E_\mathrm{F}$) at finite temperature T is $D(E)f(E)$. Hence the total number of thermally created electrons is 
\begin{equation}
n_e(T) = \int_{E = 0}^{\infty}  D(E) f(E) dE.
\label{eq:n}
\end{equation}
Here $E_F$ is taken as the reference level, $f(E)=1/(1+\exp (-E/k_\mathrm{B}T))$ is the Fermi function. In a similar manner, the total number of thermally created holes can be found by the expression
\begin{equation}
n_h(T) = \int_{E = -\infty}^{0}  D(E) [1-f(E)] dE.
\label{eq:p}
\end{equation} 
In an intrinsic semiconductor, at thermal equilibrium, the number of  thermal electrons is equal to the number of created holes i.e., $n_{ei} = n_{hi} = \sqrt{n_e n_h}$.  In addition to the thermally created charge carriers, there exists finite number of charge carriers $n_{e0}$ even at $T=0$. So the total number of intrinsic carriers at a given $T$ is $n = 2\sqrt{n_e n_h} + n_{e0}$. To obtain spin resolved total carriers one has to replace $D(E)$ with spin resolved density of states. The intrinsic spin polarization is obtained by the following expression
\begin{equation}
P(T) = \dfrac{n_{\uparrow}(T)-n_{\downarrow}(T)}{n_{\uparrow}(T) + n_{\downarrow}(T)} \times 100.
\end{equation}

\section{Experimental Results}
\subsection{Crystal Structure}

\begin{figure}[h!]
	\centering 
	\includegraphics[scale=0.33]{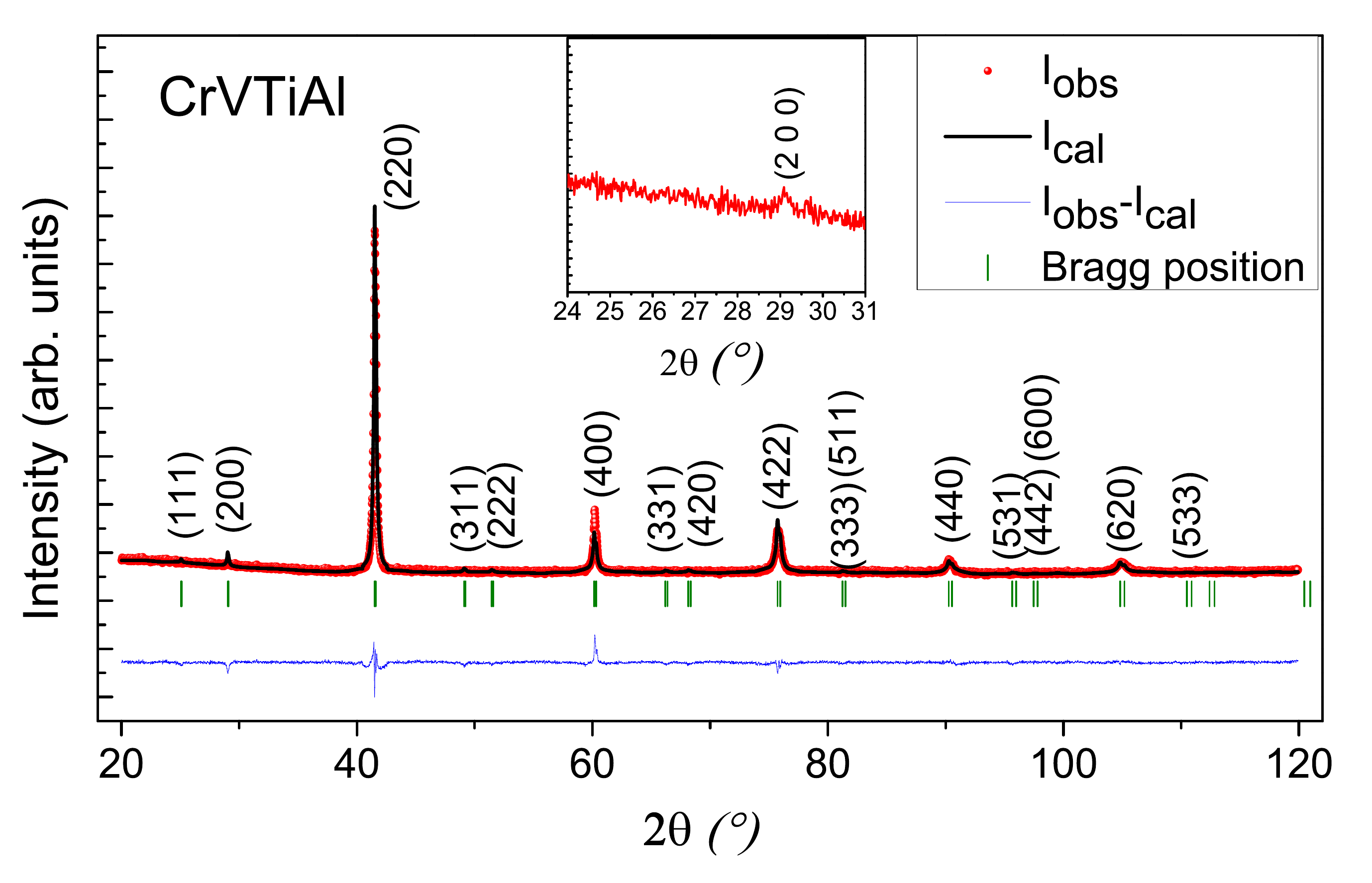}
	\caption{Rietveld refined room temperature XRD of CVTA. The super-lattice (111) reflection is absent whereas a weak (200) peak is present (see inset).}
	\label{fig:xrd-crvtial}
\end{figure}

CVTA is found to crystallize in the LiMgPdSn (space group $\mathrm{F\bar{4}3m}$, \# 216) prototype structure (or  Y  structure) with a lattice parameter of $a_\mathrm{exp}$ = 6.14 \AA. Figure 2 shows the XRD pattern for CVTA. Inset shows the zoomed region near (200) peaks. The absence of (111) and the presence of (200) superlattice peaks usually indicate the existence of B2 type disorder. For this type of antisite disorder to occur, there should be simultaneous disorder between two pairs of atoms occupying octahedral sites and tetrahedral sites i.e., disorder between one pair of X and X$'$ and another pair of Y and Z. Because of this, the resulting structure resembles the CsCl type structure. Rietveld refinement with B2 disorder did not fit well for (200) peak. Observed peak intensity was much less than the calculated value for this peak. Subsequently it was fitted to $\mathrm{DO_3}$ type anti-site disorder which yielded good agreement between the experimental data and the theoretical pattern. As Cr, V and Al have nearly same electronegativity values, it is  more probable to have anti-site disorder among these atoms. Due to the large ionic radius and least electronegativity, Ti ions are less prone to have antisite disorder with other atoms. However, it is to be noted that XRD analysis alone cannot completely resolve the structural disorder in this alloy.

\subsection{Magnetic and Transport Properties}

\begin{figure}[htbp]
\centering
\includegraphics[width=\linewidth]{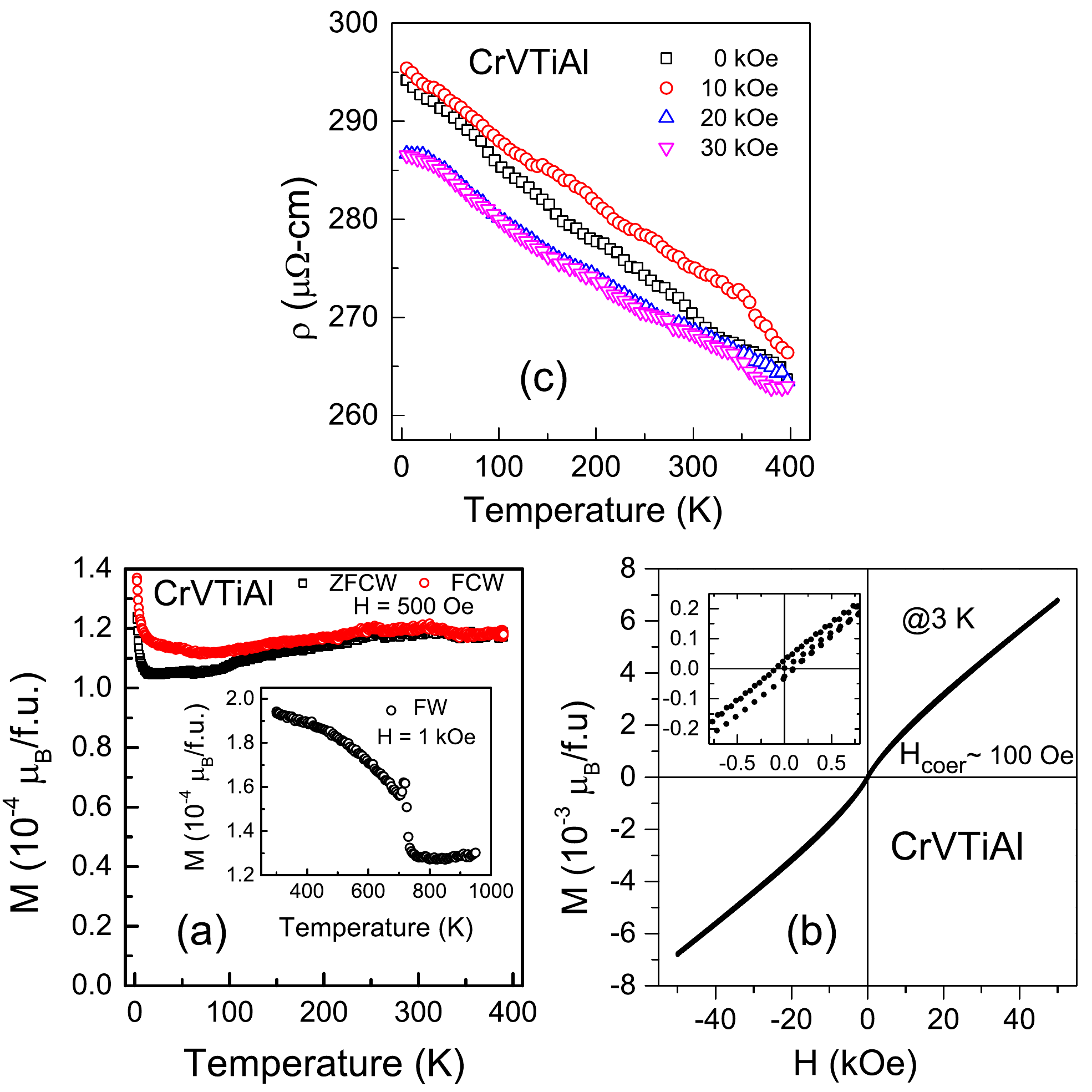}
\caption{(a) Temperature dependence of magnetization in ZFCW and FCW modes. The inset shows the temperature dependence of magnetization in high temperature regime. (b) The field dependence of magnetization. The inset shows the magnetization in zoomed scale. (c) The temperature dependence of electrical resistivity in different fields.}
\label{fig:CrVTiAl-MTandMH}
\end{figure}

For QHAs composed of atleast two elements having less than half-filled d electrons, the saturation magnetization obeys the Slater Pauling (SP) rule,\cite{k.Ozdogan-allQHA,18sprule,18sprule2}
\begin{equation}
M = N -18 \ \ \ \mu_\mathrm{B}/\mathrm{f.u.}
\label{eq.m18}
\end{equation}
where $N$ is the total number of valence electrons in the alloy.

Figure \ref{fig:CrVTiAl-MTandMH} (a) and (b) show the M-T and M-H data which clearly indicate a very small moment of CVTA ($\sim 10^{-3}\ \mu_\mathrm{B}/\mathrm{f.u.}$) and the magnetic ordering temperature is high ($\sim$ 710 K, see the inset of Fig. \ref{fig:CrVTiAl-MTandMH}(a)).  M-H curve, as shown in Fig. \ref{fig:CrVTiAl-MTandMH}(b), has a low, nonzero hysteresis (see also the inset). The behaviour remains almost the same at high T. Observation of extremely low moment with finite hysteresis indicate the strong possibility of fully compensated ferrimagnetic nature, as also found in our simulation. The outcome of nearly zero moment is consistent with the SP rule, which is a prerequisite for spintronics materials. Stephen \textit{et. al.}, \cite{expcrvtialpaper} on the other hand, reported a linear M-H curve, which may be due to the presence of small impurities present in their sample.

Figure \ref{fig:CrVTiAl-MTandMH}(c) shows temperature dependence of resistivity in different fields. The resistivity shows negative temperature coefficient, suggesting semiconducting behaviour. In intrinsic semiconductors, the variation of $\rho$ with T is dominated by exponential term, as shown in Eq. \ref{eq:intrinsic_density}. Hence the absence of such a term in CVTA indicates either gapless or spin-gapless semiconducting nature.  

\section{Theoretical Results}

We have fully relaxed the experimentally formed crystal structure (space group \# 216) in three distinct configurations, I, II and III, as described in Sec. II. The whole idea of performing these simulations was to get a better understanding of the XRD results (e.g. absence of (111) peak) which is not enough to clarify a few of the structural aspects. Table I shows the relaxed lattice parameter ($a_0$), total and atom projected magnetic moments and the total energy ($E_0$) for the three configurations.
\begin{table}
\centering
\caption{Relaxed lattice parameter ($a_0$), atom-projected magnetic moments, total moments ($\mu_\mathrm{B}$) and total energy ($E_0$) for the three configurations I, II and III of CVTA.}
\begin{tabular}{l c c c c c c}
\hline \hline
& & & & & & \\
Type& $\ $ $a_0$ (\AA) $ \ $  &  $m^{\mathrm{Cr}}$ & $\ $ $m^{\mathrm{V}}$ $\ $  &  $m^{\mathrm{Ti}}$  & $\ $ $m^{\mathrm{Total}}$ $ \ $ & $E_0$(Ry/atom) \\ & & & & & &\\ \hline \\
I    & 6.08   & 0.00 	&     0.00  	&	0.00 	& 0.00 		& -171.370934   \\ & & & & & & \\
II   & 6.15   & 2.25 	& 	-1.26   	& 	-0.98	& 0.00		& -171.370997   \\ & & & & & & \\
III  & 6.19   & 2.80	& 	-2.29 		& 	-0.49 	& 0.00 		& -171.371658   \\ & & & & & & \\
\hline \hline
\end{tabular}
\label{tab:magbeh-theor-crvtial}
\end{table}

Among these, the configuration III was found to be energetically the most stable one with lattice parameter $a_0 = 6.19$ \AA. The total energy difference among the three configurations is less than 1 $\mathrm{mRy/atom}$ which hints that at finite temperature CVTA could be a mixture of these three configurations, responsible for the observed disorder. In order to understand the effect of this disorder, we studied all the three configurations in detail.

\begin{figure}[htbp]
	\includegraphics[width=\linewidth, height=\linewidth]{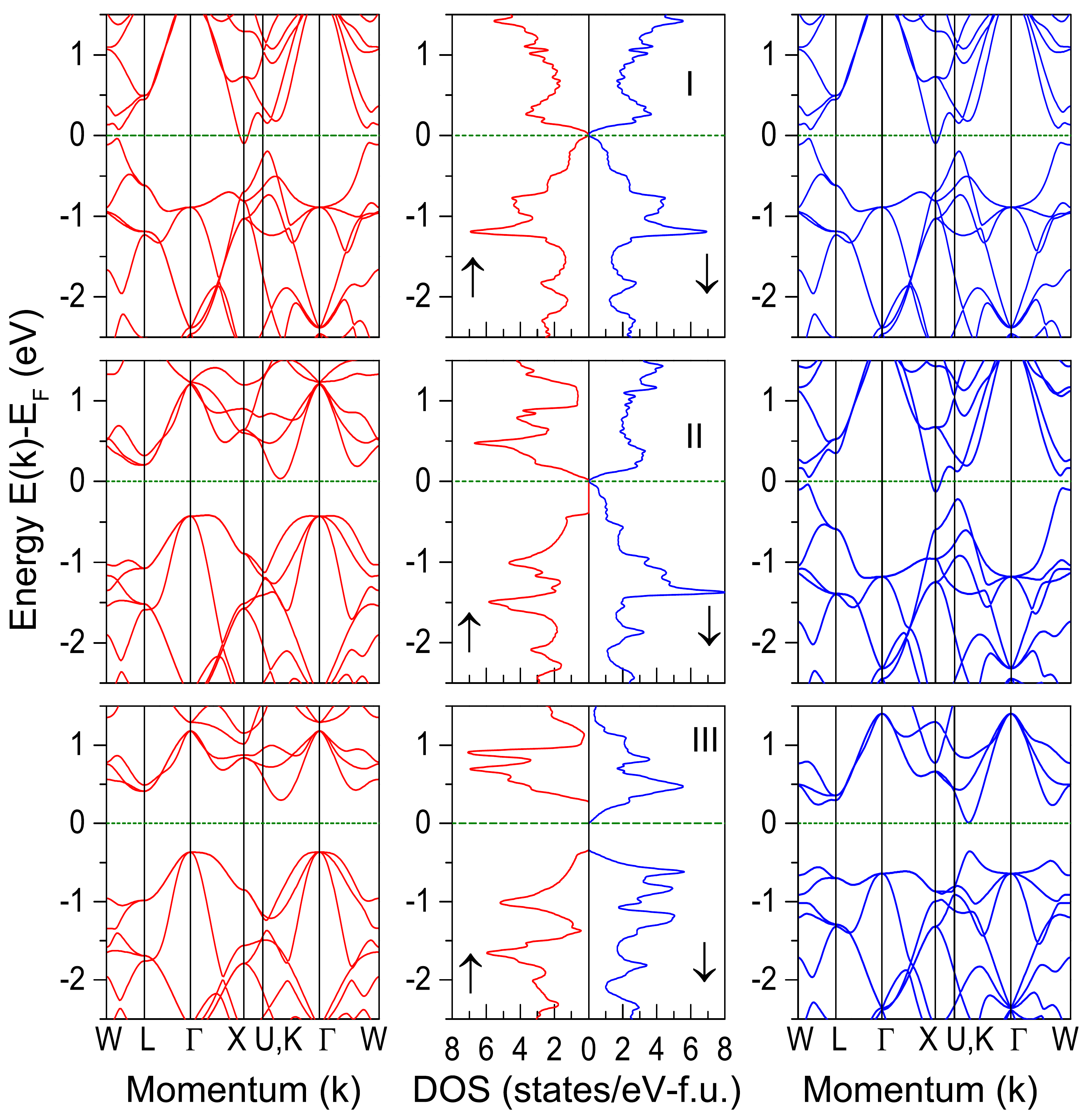}
	\caption{Spin polarized band structure and DOS for the configurations, I, II and III of CVTA at relaxed lattice parameters ($a_0$). Left-side bands correspond to spin up while right-hand side is for spin down.}
	\label{fig:DOSBS-all}
\end{figure}

\begin{figure}[hbtp]
\includegraphics[width=\linewidth]{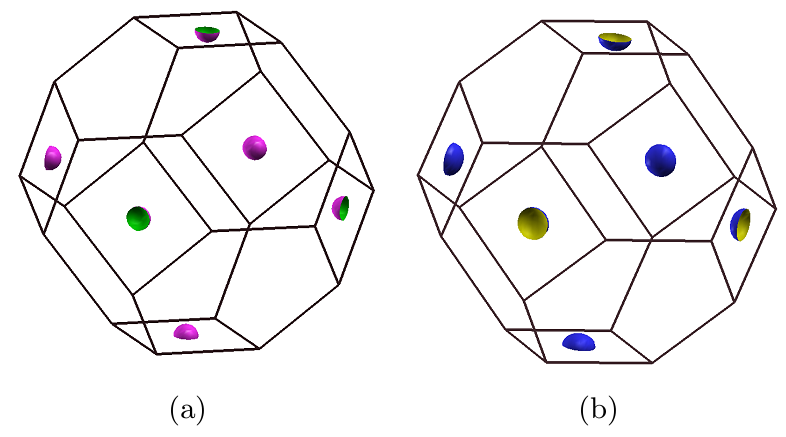} 
	\caption{(a) Fermi surface (FS) is same for both up and down spin bands of configuration I. It has spherical shape and crosses at X point. (b) Fermi surface for the spin down band of configuration II. It has oblate shape and crosses at X point. Configuration III does not have any FS. }
	\label{fig:fermi-CrVTiAl}
\end{figure}

Calculated spin polarized band structure and DOS for all the three configurations are shown in Fig. \ref{fig:DOSBS-all}. Calculations for all the three configurations were initiated with a ferrimagnetic arrangement of spins with moments at Cr-atoms aligned antiparallel to those of V and Ti. This was done keeping in mind the vanishingly small experimental net moment (see previous section) and other theoretical reports where ferrimagnetic arrangement was proposed to be the stable phase. In our case, configuration I converges to a non-magnetic phase with identical spin up and down bands, and consequently has the lowest magnetic ordering temperature. Both spin up and down bands are gapless with nearly zero DOS at $E_\mathrm{F}$, indicating the gapless nature. It acquires an indirect band gap with conduction band minima touching at X-point and valence band maxima at other $k$-point. Figure \ref{fig:fermi-CrVTiAl}(a) shows the Fermi surface plot for configuration I. As expected, both spin up and down Fermi surfaces are identical, with tiny spherical shape and are shared by neighbouring Brillouin zone. The essential features of DOS and band structure remain unchanged at $a_{\mathrm{exp}}$ and hence its physical properties (transport and magnetic) are robust.

In the case of configuration II, irrespective of the initial magnetic moments at each site, the calculations converged in a ferrimagnetic arrangement with Cr moments aligned antiparallel to V and Ti. For this configuration, DOS and band structure for spin down channel mostly resemble that of configuration I, except for the shape and size of Fermi surface at X-point, indicating gapless nature for spin down channel. The shape of the Fermi surface, as shown in Fig. \ref{fig:fermi-CrVTiAl}(b) is oblate, centred at $X$-point and equally shared by neighbouring Brillouin zone. The size of the surface is more than double that of configuration I. There is almost no change in the DOS and band structure of spin down channel with $a_0$ and $a_\mathrm{exp}$, indicating that its spin down gapless nature is robust against small changes in $a$. On the other hand, DOS and band structure for spin up channel shows a clear indirect band gap of $\Delta E_g^\uparrow = 0.36$ eV, revealing semiconducting nature. There is no observable change in $\Delta E_g^\uparrow$ with lattice parameter indicating that its semiconducting nature is also robust against small changes in $a$. Due to the absence of symmetric DOS and band structure along with relatively high absolute magnetization (4.64 $\mu_\mathrm{B}/\mathrm{f.u.}$), its magnetic ordering temperature is expected to be very high. Such a phase with zero gap in one spin band and finite gap in the other gives rise to a fully compensated ferrimagnetic, spin-gapless semiconductor.

Similar to configuration II, configuration III is also found to be ferrimagnetic. However the later has $\Delta E_g^\uparrow = 0.58$ eV and $\Delta E_g^\downarrow = 0.30$ eV at $a_0$, indicating that it is a magnetic semiconductor. Spin up and down gaps are reduced to 0.55 eV and 0.25 eV respectively at $a_\mathrm{exp}$. Absence of Fermi surfaces (no states at $E_\mathrm{F}$) for both spin channels also confirms the magnetic semiconducting nature. Presence of large exchange splitting gaps and a large absolute magnetization (5.92 $\mu_\mathrm{B}/\mathrm{f.u.}$) indicates a large magnetic ordering temperature. 

All the properties of configuration III (such as magnetic state, $\Delta E_g^{\uparrow,\downarrow}$ etc.), which corresponds to ground state one, are in good agreement with the earlier reports by Ozdogan \textit{et. al} with exception of sign of moments on all ions. As a result, DOS and band structure are interchanged for spin up and down electrons. In addition, they reported a small positive moment on Al ions which also has an opposite sign in the current study. Notably, Cr and V moments are aligned in opposite directions and the resulting moment is compensated by Ti ions (see Table \ref{tab:magbeh-theor-crvtial}).

\begin{figure*}[tbhp]
	\includegraphics[width=\linewidth, height=0.30\linewidth]{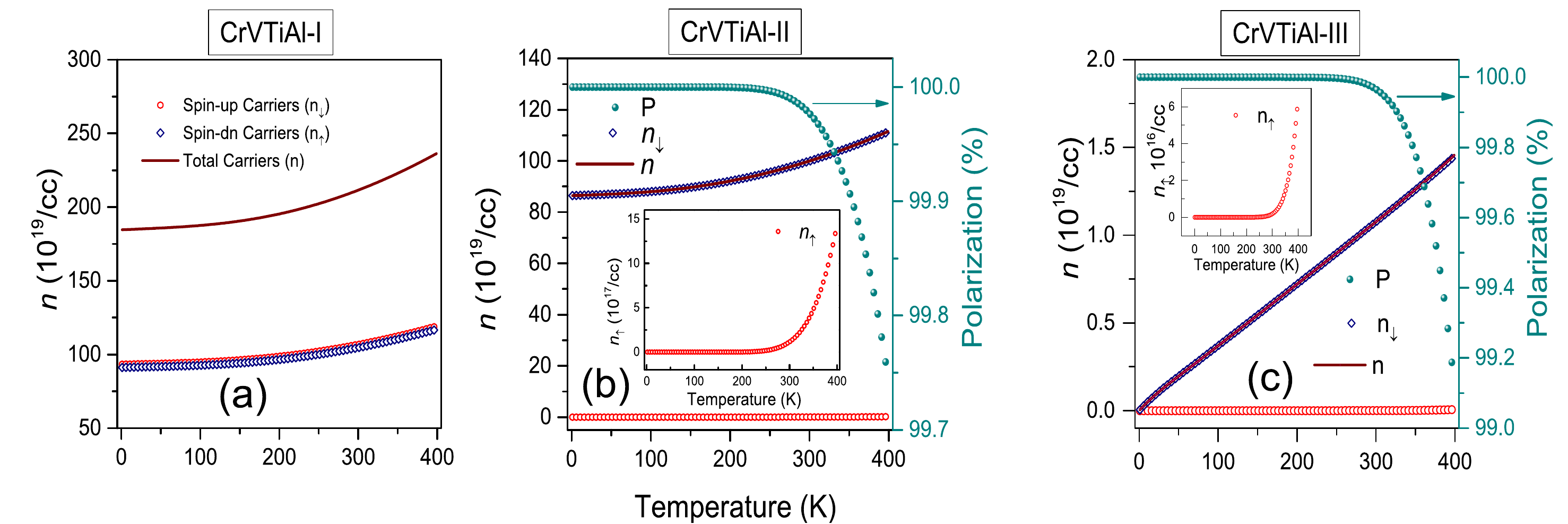}
	\caption{Spin resolved intrinsic carrier concentrations (left axis) and spin polarization (right axis) vs. temperature (between 3-400 K) for configuration I (left), II (middle) and III (right panel) for CVTA.}
	\label{fig:carrierconcentration}
\end{figure*}

It is important to note that all the configurations show different forms of semiconductors, with two of them converged into the ferrimagnetic state. In order to understand the true nature of the semiconductor, it is quite relevant to study the temperature dependence of the transport properties i.e., intrinsic carrier concentration ($n$) and spin polarization ($P$). Figure \ref{fig:carrierconcentration} shows the temperature dependence of these quantities for the three configurations. Configuration I, being almost non magnetic, has negligible spin polarization ($P$), although they indeed have a finite carrier concentration due to small states at the Fermi level ($E_\mathrm{F}$), For configuration II and III, the spin up carrier concentration is negligibly small due to vanishing states at $E_\mathrm{F}$. The spin down carrier concentration, on the other hand, is large but has very different nature of T-dependence for the two configurations (II and III). Interestingly, in the case of configuration III, $n$ shows an almost straight line behaviour, which is neither exponential nor T$^{3/2}$. The magnitude, however, is much smaller ($n \sim 2$) compared to the other two configurations. Spin polarization values for these two configurations are high.

\section{Discussion and Conclusion}

Experimental results reveal that CrVTiAl is a fully compensated ferrimagnet with $\mathrm{DO_3}$ disorder among $\mathrm{Al}$, $\mathrm{Cr}$ and V atoms. First principle calculations within GGA approximation predict configuration III as the ground state, which is a fully compensated ferrimagnet with unequal band gaps for spin up and down channels. However, the energy differences among all the three configurations (I, II and III) are small, which indicates the possibility of a mixed phase or disordered phase at finite T. One can consider the disordered state as a linear combination of all the three independent configurations (with probabilities proportional to their Boltzmann factors) yielding either spin-gapless, gapless semiconductor or magnetic semiconductor with reduced band gaps. This concept originates from the fact that the disordered Kohn Sham (KS) orbitals ($\Phi_{dis}$) can be written as linear combination of  KS orbitals ($\Phi$) of each configuration i.e., $\Phi_\mathrm{dis}= c_\mathrm{I} \Phi_\mathrm{I} + c_\mathrm{II} \Phi_\mathrm{II} + c_\mathrm{III} \Phi_\mathrm{III} $ where $c_i\propto \exp({-E_0^i/{k_\mathrm{B}T}})$. Such a linear combination is possible due to the fact that all the three pure configurations have vanishing states at the Fermi level, unlike most of the reported cases in which certain configurations alone have finite states at $E_\mathrm{F}$. At the observation level, the above linear combination presents itself as having a predominantly SGS property, because the first term is energetically least attainable and non magnetic while the third term is not effective as the DOS ($D_{\uparrow,\downarrow}(E)$) is zero at the Fermi level ($E_g^{\uparrow,\downarrow}\gg k_{\mathrm{B}}T$). The last scenario can change if there are impurities, which will alter the SGS nature seen in our sample.\cite{expcrvtialpaper}

In conclusion, we identify the true crystallographic and magnetic ground states of CrVTiAl with the help of a joint theoretical and experimental investigation. While the magnetic ground state is uniquely identified as a fully compensated ferrimagnet, the balance between spin gapless nature and the magnetic semiconducting nature appears to be quite delicate  according to the theoretical calculations. However, by combining the experimental data of transport measurement, we are able to conclude that the alloy is predominantly a spin gapless semiconductor. 


\section*{Acknowledgements}
Y. Venkateswara acknowledges the financial help provided by IIT Bombay for carrying out this research. A. Alam acknowledge DST-SERB (SB/FTP/PS-153/2013) for funding to support this research. S. S. Samatham and Enamullah acknowledge IIT Bombay for the financial support through Institute Post Doctoral Fellowship.


\bibliography{bib}

\end{document}